\input harvmac.tex
\input epsf

\def\np#1#2#3{Nucl. Phys. {\bf B#1} (#2) #3}



\def\U{{\cal U}}


\line{\hfill PUPT-1709 }
\line{\hfill {\tt hep-th/yymmddd}}
\vskip 1cm

\Title{}{From 0-Branes to Torons.}

\centerline{$\quad$ { Z. Guralnik and S. Ramgoolam}}
\smallskip
\centerline{{\sl Joseph Henry Laboratories}}
\centerline{{\sl Princeton University}}
\centerline{{\sl Princeton, NJ 08544, U.S.A.}}
\centerline{{\tt zack,ramgoola@puhep1.princeton.edu}}

\vskip .3in

The moduli space of 0-branes  on $T^4$ gives a prediction for 
the moduli space of torons in $U(n)$ Super Yang Mills theory  
which preserve $16$ supersymmetries.  The zero brane number 
corresponds to the greatest common denominator of the rank $n$, 
magnetic fluxes and the instanton number.  This prediction is 
derived using U-duality.  We explicitly check this prediction 
by analyzing $U(n)$ bundles with non-zero first as well as second 
Chern classes.  The 
argument is extended to deduce the moduli space of torons which 
preserve $8$ supersymmetries. Parts of the discussion  extend 
naturally to $T^2$ and $T^3$.  Some of the  U-dualities 
involved are related to Lorentz boosts along the eleventh 
direction in M theory.


\Date{8/97}

\lref\bfss{ T. Banks, W. Fischler, S.H. Shenker and L. Susskind,
  {\it ``M Theory As A Matrix Model: A Conjecture,''} 
 Phys.Rev.D55:5112-5128,1997. hep-th 9610043}
\lref\dlp{ J. Dai, R. Leigh, J. Polchinski, 
{\it ``New connections between string theories,''},  
Mod. Phys. Lett. A4, 2073, 1989  }
\lref\pol{ J. Polchinski,{\it ``Combinatorics 
of boundaries in string theory, ''} Phys.Rev.D50:6041-6045,1994 }
\lref\hor{P. Horava, {\it 
``Strings on world-sheet orbifolds,''}, Nucl.Phys.B327:461,1989 } 
\lref\green{M. Green, {\it ``Point-like states for type IIB superstrings,''} 
 Phys.Lett.B329:435-443,1994 }
\lref\hvtalk{H. Verlinde, talk given at Princeton,  Nov. 1996}
\lref\vjl{ V. Balasubramanian and R. Leigh, { \it ``D-branes, 
moduli and supersymmetry, '' } Phys. Rev. D55, 6415, 1997}
\lref\dtor{ Z. Guralnik and S. Ramgoolam, 
``{\it Torons and D-brane bound states,} '' hep-th 9702099, 
  Nucl. Phys. B499:241,1997} 
\lref\hatay{A. Hashimoto, W. Taylor, ``{\it Fluctuation spectra 
of tilted and intersecting D-branes from the Born-Infeld action}'' 
hep-th 9703217 } 
\lref\tseyt{ A. Tseytlin, 
``{\it On non-abelian generalisation of Born-Infeld action
 in string theory,}'' hep-th 9702163}
\lref\berdole{ Berkooz, Douglas, and Leigh,
 ``{\it Branes intersecting at angles,}'' Nucl. Phys B480 (1996) 265-278 }
\lref\gnt{ E. Gava, K.S.Narain, M.H.Sarmadi, ``{ \it On the bound states 
    of p-and (p+2)-branes}.'' hep-th 9704006. }
\lref\tay{ 
W. Taylor, ``{\it Adhering 0-branes to 6-branes and 8-branes,}''
 hep-th/9705116. } 
\lref\dnotes{ J. Polchinski, S. Chaudhuri, C. Johnson, 
``{\it Notes on D-branes,}''  hep-th 9602052  }
\lref\hamo{ J. Harvey and G. Moore,{\it ``On algebras of BPS
states, '' hep-th 9609017  }  } 
\lref\bsv{ M. Bershadsky, V. Sadov, C. Vafa, \np {463}{1996}{420} }

\lref\dvv{ Dijkgraaf, Verlinde, and Verlinde, {\it Matrix strings} }
\lref\thft{ t'Hooft, ``{\it Some twisted self dual solutions of the 
			    Yang-Mills equations on a hypertorus,}''
 Commun Math. Phys.81, (1981) 267-275. }
\lref\hms{ J. Harvey, G. Moore, and A. Strominger, ``{\it Reducing 
           S-duality to T-duality,}'' Phys.Rev.D52 (1995) 7161-7167, 
           hep-th 9501022. } 
\lref\vbal{ P. van Baal,{\it  ``Some results for SU(N) gauge 
fields on the Hypertorus,''}, Comm. Math. Phys. 85, 529 (1982) } 
\lref\ed{ E. Witten ``{\it Constraints on supersymmetry breaking,}''
 Nucl. Phys. B202, (1982) 253-316. }
\lref\giv{A. Giveon, M. Porrati, E. Rabinovici,{\it ``Target space duality
in string theory,'' } Phys. Rept. 244:77-202, 1994} 
\lref\witbd{ E. Witten, { \it ``Bound states of strings and P-branes}, 
    Nucl. Phys. B460: 335-350, 1996, hep-th/9510135. }
 \lref\hv{F. Hacquebord, H. Verlinde, 
{\it ``Duality symmetry of $N=4$ Yang Mills on $T^3$ }, hep-th/97070179.  }
\lref\ls{ L. Susskind, { \it ``Another conjecture about M(atrix) Theory}, 
                  hep-th 9704080.  }  
\lref\gop{ R. Gopakumar, { \it ``BPS states in Matrix strings,''} 
               hep-th/9704030}                  
\lref\cjp{S. Chaudhuri, C. Johnson,  J. Polchinski, 
          {\it ``Notes on D-branes''} hep-th/9602052 . }
\lref\schw{ J. Schwarz, {\it ``The power of M theory,''
Phys.Lett. B367 (1996) 97-103, hep-th 9510086 }    } 
\lref\asp{ P. Aspinwall, {\it ``Some relationships between dualities 
 in string theory,''},      
Nucl.Phys.Proc.Suppl. 46 (1996) 30-38,  hep-th/9508154.  } 
\lref\nish{H. Nishino, {\it `` Supersymmetric Yang Mills theories 
in $D \ge 12$''} hep-th/9708064. } 
\lref\mm{ A.Losev, G. Moore, S. Shatashvili, {\it ``M\&m's'' }, 
hep-th/9707250.  }
\lref\vafa{ C. Vafa, {\it ``Instantons on D-branes,''} 
            Nucl.Phys. B463 (1996) 435-442, hepth/9512078 } 
\lref\holiwu{ M. Li, P.M.Ho, Y.S.Wu, {\it ``$p-p^{\prime}$ strings 
                           in Matrix Theory,''} hep-th/9706073. }  
\lref\domo{ M. Douglas, G. Moore, 
{ \it ``D-branes, quivers and ALE instantons,''} hep-th 9603167. } 
\lref\grt{ O. Ganor, S. Ramgoolam, W. Taylor,
 {\it ``Branes, Fluxes and Duality in Matrix theory,''} hep-th/9611202, 
 Nucl. Phys. B492, 191-204, 1997 }
\lref\bss{ T. Banks, N. Seiberg, S. Shenker {\it ``Branes from Matrices,''}
             hep-th/9612157, Nucl. Phys. B490;91-106, 1997 } 
 \lref\hamada{K. Hamada, {\it ``Vertex operators for SuperYAng-Mills
 and Multi-D-branes in Green-Schwarz Superstring,''} hep-th/9612234,
      Nucl. Phys. B497: 511-524, 1997 } 
\lref\bkop{E. Bergshoeff, R. Kallosh, T. Ortin, G. Papadopoulos, 
           {\it ``Kappa Symmetry, Supersymmetry, and Intersecting Branes,''}, 
            hep-th/9705040 }  
\lref\gil{ G. Lifschytz, {\it ``four-brane and six-brane interactions 
                                in MAtrix theory,''}, hepth-9612223  } 
\lref\witsma{ E. Witten, {\it ``Small instantons in string theory,''} 
                        Nucl. Phys. B460:541-559, 1996,
 hep-th/9511030. }          
\lref\doug{M. Douglas, {\it ``Branes within branes,''}, hep-th/9512077.}

\newsec{ Introduction} 

The connection between D-branes
\refs{ \dlp, \pol,  \hor, \green } 
 and torons \thft\
was studied in  some detail \refs{\dtor, \berdole,
\vjl,\hatay, \gnt, \holiwu }. Torons also appeared in the 
 context of Matrix theory in \refs{ \grt, \bss}.   
Torons are related to systems of branes at angles, 
where the angles involved are mapped to field strengths. 
In this paper we show how 
 some detailed  properties of moduli spaces of 
torons  follow from their connection to D-brane
bound states.

Section 2 discusses most of the key points in the simpler 
case of $U(N)$ bundles on $T^2$. 
For simplicity we will restrict attention to tori
which have diagonal metric. 
We first describe  how dualities 
can be used to map a system of $N$ 2-branes 
with $m$ zero branes to a system of D-strings
at an angle. The map to D-strings gives a hint about
the moduli space, but a more precise prediction 
is obtained by mapping to zero branes. 
The basic strategy is to map via dualities the system 
of 2-brane and 0-brane related to torons, to a system where
the moduli space of supersymmetric solutions is known and has 
a simple geometrical interpretation. The counting of
BPS states is invariant under the dualities. Further, 
since these are systems with a lot of supersymmetry
the bound state spectrum is determined largely by the 
properties of the moduli space. Hence we can expect the moduli 
space to stay unchanged. These ideas were used to motivate
Nahm duality exchanging instanton number and rank of gauge group
for $T^4$ bundles by H. Verlinde \hvtalk, and also appeared in \domo.  
After briefly describing 
the dualities involved, we do the gauge theory 
calculation which establishes that the moduli space is as expected.

In section 3, we recall the setup of 
\dtor, where torons ( constant field strength solutions
on $T^4$)  were related to systems of 4-branes, 2-branes and zero-branes. 
The relation to systems of 2-branes at angles 
described in \dtor\ gives a hint about the structure
of the moduli space. Again more information is obtained 
by mapping to zero branes.  
In terms of these,   the moduli space is 
entirely geometrical, given by the coordinates of  $p$ 0-branes, 
where $p $ is related to the rank and fluxes as described in 
detail in section 3.
Therefore the dimension of the moduli space of gauge inequivalent
torons with the same field strengths should be $(T^4)^p/S_p$.  
This is equal to the moduli space of flat connections,  since
the torons which
preserve $16$  supersymmetries have non-zero $Tr F_{\mu\nu}$,  
but vanishing $SU(N)$ field strength.
We explicitly count the 
number of flat connections of the gauge potential which survive
when such twisted boundary conditions are imposed,  and
verify that there are $4p$ independant zero modes. 
The toron solutions which are dual to 0-branes are particularly simple
in the sense that the $SU(N)$ field strengths vanish and half the 
supersymmetries are preserved.  However by simple arguments 
discussed in  \dtor,   we extend these results to 
more general toron solutions 
associated with intersecting 2-branes which  
preserve a quarter of the supersymmetries.

In section 4, we extend the discussion to 
 class of solutions on $T^3$ as an example of 
how facts about moduli spaces  on odd dimensional tori 
 can also be obtained from dualities.

\newsec{  Torons on $T^2$. }

\subsec{ Prediction from dualities} 
For simplicity,  we begin by considering the moduli space 
of the torons on $T^2$ in maximally supersymmetric 
 $U(N)$ Yang Mills on $T^2 \times R$. 
We will predict    the moduli space
by mapping to a system of D-strings. We also describe a map 
to zero branes which makes the moduli space 
that of positions of the zero branes.   Then we  will
explicitly determine the space of flat connections in a twisted
sector of the Yang-Mills theory and find that it matches.

Consider $N$ parallel 2-branes wrapped on $T^2$ bound to $m$
0-branes.  This corresponds to a $U(N)$ Yang-Mills theory in 
a twisted sector with magnetic flux $m$.  
By a T-duality in the $1$ direction,   we obtain D-strings with 
charges $Q_1 = m$ and $Q_2= N$. 
When the greatest common denominator,  $ g.c.d (N,m) =p$  
is one, the D-string wraps once around a cycle 
which is a linear combination of $N$ times one generating cycle
and $m$ times another cycle. So we expect a moduli space 
for the position of the D-string center of mass on $T^2$. 
When $p$ is greater than one, we can have a family of 
flat connections on the D-string worldvolume, some 
of which describe winding  configurations of the 
D-strings, but most of which have no geometrical interpretation 
\cjp. 
  This is illustrated
in Fig.1.  The various wrappings are characterized by
holonomies of the D-string effective action which form the Weyl subgroup
of $SU(p)$.  The rest of the moduli space interpolates between the
wrapping geometries.
This suggests that the moduli space should be 
a $U(p)$ moduli space of flat connections on $T^2$. 
This is a $2p$ dimensional space $(T^2)^p/S_p$,
$p$ of which is roughly  related to the wrapping geometries, and 
the other $p$ is roughly related to positions of D-strings. 
If we map to zero branes then the moduli 
space is entirely given by positions of the $p$ zero branes. 

There is an element of the U-duality group 
which acts on  the charges $(N,m)$ of a  system 
of $N$ two-branes and $m$ zero-branes, by shifting 
$m$ by a multiple of $N$. One way to see this is to compactify 
on a further circle, do a T-duality to convert the $(20)$ system 
to a $(31)$ system, then use S-duality to get 
3-brane and elementary string. Then do further two 
T-dualities to get D-string and elementary string. Here 
we know from \witbd\ that we can shift the
elementary string charge by a multiple of the D-string charge
( or vice versa). We can then transform back to the 
$(02)$ system with shifted charges. Using these symmetries we can 
map the $(N,m)$ system to a $(0,p)$ system, i.e a system with 
zero brane charge only. 
Another way to see that one can get pure  zero branes is 
to consider the T-duality group $O(2,2,Z)$,  in which 
case we do not have to compactify on an extra circle. 
The theta-shift symmetry \giv\ which changes  the  combination 
of $U(1)$ gauge field strength and NS 2-form,   $(F+B)$, 
by $1$, and hence the zero brane charge of 
$N$ two-branes by $N$, can be combined with the usual 
factorized T-dualities to convert the $(N,m)$ system to 
a $(0,p)$ system. Since the theta shift is of the 
form $T_1ST_1$ where $T_1$ is a factorized T-duality
and $S$ is in  the geometrical  $SL(2,Z)$, the above element
is  a product of a transformation 
that maps the $(N,m)$ system to D-string at an angle 
with a T-duality at an angle. Note that this generalizes Nahm duality
which exchanges $N$ and $m$.  to an $SL(2,Z)$ action 
on the pair $(N,m)$.
 For $p$ zero branes on $T^2$,  the moduli space is 
$(T^2)^p/S_p$.  Since in these highly supersymmetric systems 
we can expect the moduli space to be invariant under the 
dualities, we have a prediction for the moduli space 
$U(N)$ with flux $m$.

 The shift of the zero brane charge 
 of a system of zero and 2-brane charge, by a multiple of the 
2-brane charge,  has a simple interpretation in terms of the 
picture of type IIA
string theory  on $T^2$ as M theory compactified 
on $T^3$. It is due to Lorentz invariance. 
Consider for example $N$ two-branes. 
Adding $Nk$ units of zero brane charge 
amounts to boosting by $Nk$ units of momentum along the eleventh 
direction. We can 
partition the momentum  equally between the 
$N$ two-branes, with each having $k$ units.
This defines a unique map from the system of 
bound states ( including multi-particle states)
 of charge $(N,0)$ to those of  charge $(N,kN)$.   
The $(N,0)$ system is therefore related by 
eleven dimensional Lorentz  
symmetry  
to the $(N,kN)$ system.   (This is analogous to how 
one interprets $SL(2,Z)$ 
of type IIB in terms of reparametrizations of 
the two-torus of M theory \schw\asp).  
If we start with 
a single bound state of two-branes we can boost
it by any number of units to get another state in the Hilbert space, 
but when the boost is  a multiple of $N$ we can expect 
a symmetry of the multi-particle spectrum.

\subsec{ Gauge theory calculation} 

Now let us return to the original 
system of $N$ parallel 2-branes
described at by a $U(N)$
Yang Mills theory.  This system will also have a zero brane charge if
$Tr ( F_{12} + B_{12} )$ is non-vanishing.  
The gauge potentials of the $U(N)$ theory on a torus satisfy 
boundary conditions
\eqn\twist{A(x+a_{\mu}) = \U_{\mu}(x)(i\del + A(x))\U^{\dagger}_{\mu} (x)}
where $a_{\mu}$ are cycles around the torus.  
For a well defined bundle,  
\eqn\twisted{\U_{\mu} (x)     \U_{\nu} (x+a_{\mu})  
		\U^{\dagger}_{\mu} (x+a_{\nu})
		\U^{\dagger}_{\nu} (x) = 1,  }
in the presence of zero $B$ \foot{If $B \ne 0$, the right 
hand side is modified to  $e^{i\int B }$.}. 
$\U_{\mu}$ may be written as a product of a 
$U(1)$ part and an $SU(N)$ part
with the $SU(N)$ part satisfying 
\eqn\part{U_{\mu} (x)     U_{\nu} (x+a_{\mu})  
		U^{\dagger}_{\mu} (x+a_{\nu})
		U^{\dagger}_{\nu} (x) = e^{2\pi i {n_{\mu\nu}\over N}}}
with antisymmetric integer $n_{\mu\nu}$.
In 2 dimensions the supersymmetric solutions have vanishing $SU(N)$ field
strengths,  but the twist $n_{12}$ may be nontrivial,  and is canceled by
$U(1)$ twist arising from a non vanishing $Tr F_{12}$.  
\eqn\flx{m = {1\over 2\pi} \int d^2 x tr F_{12} = n_{12}.}  

In a gauge in which the $SU(N)$ potentials are constant,  one can find
constant matrices $U_{\mu}$ satisfying \part\ by writing 
\eqn\sts{ U_{\mu} = Q^{s_{\mu}}P^{t_{\mu}},}
where \eqn\QPQP{s_{\mu}t_{\nu} - s_{\nu}t_{\mu} = n_{\mu\nu} mod N}
and 
\eqn\QPQP{PQ = QP e^{2\pi i\over N . }}
We shall choose the convenient representation
\eqn\wepick{
Q = 
\pmatrix{
1& & & \cr
& e^{2\pi i\over N} & & \cr
& & \ddots & \cr
& & & e^{2\pi i (N-1)\over N}
},\qquad
P =
\pmatrix{
 & 1 & & \cr
 &   & 1 & \cr
 &   &   & \ddots \cr
1&   &   &  
} }
A solution of \QPQP\ for flux $m$ is given by $U_1 = Q$ and $U_2 = P^m$.
The $SU(N)$ gauge potentials then satisfy the conditions
\eqn\tone{\eqalign{ 
& A_{\mu} = Q A_{\mu} Q^{-1} \cr
& A_{\mu} = P^m A_{\mu} P^{-m} \cr }} 
The first condition restricts $A_{\mu}$ to be diagonal. 
The second condition constrains this diagonal matrix to be invariant 
under 
cyclic permutations of length $m$.  The number of independent diagonal 
elements of $A^{\mu}$ is therefore $p$ where $p= gcd(N,m)$.  Since 
there are
two dimensions,  a $2p$ dimensional moduli space of flat connections
survives \foot{The elimination of $SU(N)$ torons in 
the case $m=1$, $p=1$ 
 was  used  in \ed,  where  the 
removal of zero modes facilitates the computation of $tr (-1)^F$.}, 
 which is consistent with the zero brane 
description.


\vskip2.5cm
{\centerline{\epsfsize=3.0in \hskip 0cm \epsfbox{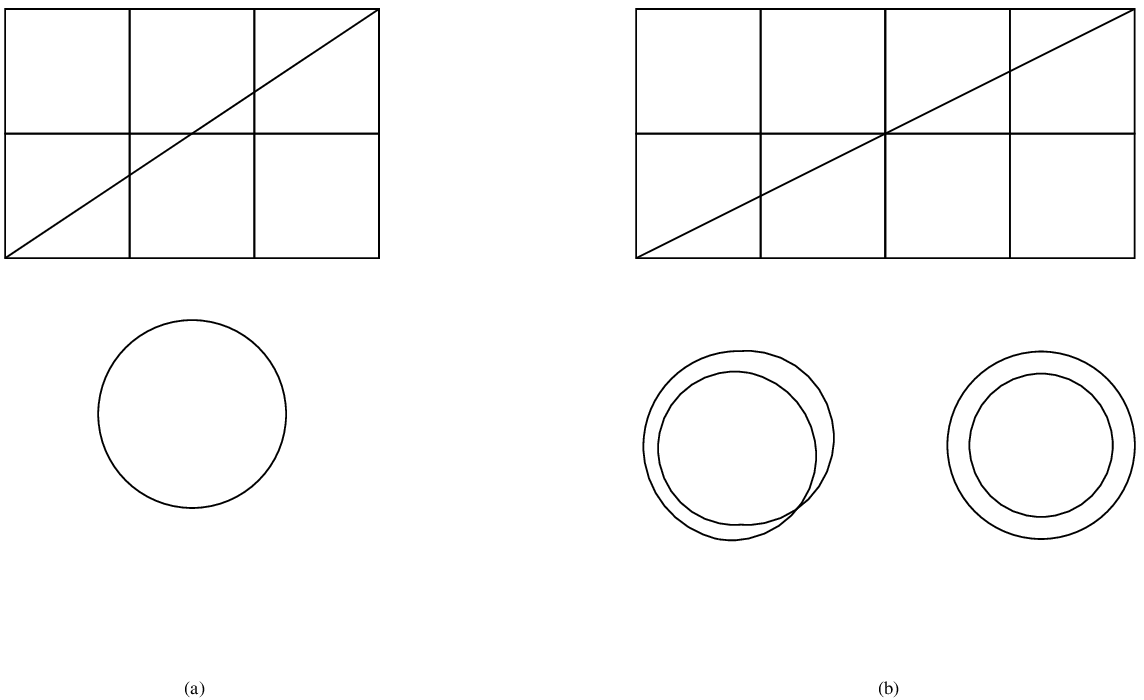}}}
\vskip0cm
{\tenrm{
FIGURE 1. \hfil\break

Two D-string configurations on $T^2$.  (a) has $N=3$ and $m=2$
so that $p=gcd(N,m)=1$,  and there is only one way to wind the string.
(b) has $N=4$ and $m=2$ so that $p=2$, and there may be two copies of
a singly wound string,  or one copy of a doubly wound string.
}}

\vskip2.0cm


\newsec{ Torons on $T^4$}

Let us recall the set-up from \dtor. We have 
maximally supersymmetric $U(N)$ Yang Mills theory in $4+1$ 
dimensions, on $T^4\times R$, 
 as the  world volume theory of $N$ 4-branes. 
First restrict attention to configurations where 
the field strengths are proportional to the 
identity in the $U(N)$ Lie algebra. They leave 
unbroken 
$16$ supersymmetries.  These were discussed 
under the heading $(4220)$  type solutions in \dtor.
The discussion of supersymmetry there was based on 
an equation in \hamo. More detailed discussions 
can be found for example in \hamada\bkop. 
These constant field strength solutions of Yang-Mills theory on 
$T^4$  correspond to BPS bound states of 4-branes, 2-branes and
0-branes.  The 0-brane charge is given by the instanton number $\nu$, 
2-brane charge by the t'Hooft twists $n_{\mu\nu}$, and 4-brane charge
by the rank of the gauge group $N$.  Using an $SL(4,Z)$ isomorphism of 
the toron solutions \vbal,  one can set all twists but 
$n_{12}, n_{34}$ and $n_{14}$ equal to zero without effecting the 
rank or the instanton number.  Then    
T-dualities in the 1 and 3 directions yield 
systems of flat 2-branes at angles,  with 2-brane charges
\eqn\chgs{\eqalign{     Q_{24} = N,\cr 
			Q_{12} = n_{14},\cr 
			Q_{34} = n_{32},\cr 
			Q_{13} = \nu, \cr
                        Q_{14} = n_{12} \cr 
			Q_{14} = Q_{23} = 0.\cr}}
If we generalize to solutions which break the $U(N)$ 
symmetry, (and preserve only eight supersymmetries), 
these charges may be written as the sums of contributions coming
from 2-branes at different angular orientations,
$Q_{\mu\nu} = \sum_i Q^{(i)}_{\mu\nu}$ where 
$\epsilon_{\mu\nu\alpha\beta}Q^{(i)}_{\mu\nu}
Q^{(i)}_{\alpha\beta} = 0$. 
Initially we will only
concern ourselves with solutions preserving $16$   supersymmetries.
All 2-branes are parallel in this case and we will drop the index $i$.
The generalization to the 
case where $i$ runs over a set of branes, 
involves solutions where the field strengths take the form 
of diagonal matrices breaking the gauge group to 
$U(k) \times U(N-k)$ say in the case where $i$ runs from $1$ to $2$. 
 Here the moduli space takes the form $(T^4)^p_1/S_{p_1} \times 
(T^4)^{p_2}/S_{p_2}$, and the proof consists in trivially extending
the proof we give below for the unbroken symmetry case
 to each unbroken 
group $U(k)$ and $U(N-k)$.  

\subsec{ Mapping the toron to zero brane}

We can  use U-duality to map 
a system of 4-branes and 2-branes preserving 
 16 supersymmetries to a system of pure zero-branes. 
 For simplicity take the $(4220)$ system which 
 is obtained from Yang Mills with fluxes 
 $n_{12}$, $n_{34}$. This is T-dual to a system 
 with 2-brane charges along $(24)$,  $(13)$, 
 $ (14)$ and $(32)$. After another T duality along 
   the $2$ axis, we get three brane charge along
   $(123)$, $(124)$ and 1-brane charge 
   $(4)$ and $(3)$. An  S-duality gives a  three
    brane and elementary string. Finally a  T-duality
  along $(1)$   and $(2)$ gives a  D-string 
  parallel to a an NS-string in the $(34)$ plane.  
 The fact that the D-string and NS-string are parallel  
  follows from the zero self-intersection number 
 of the original 2-brane system. Now we know from 
 \witbd\ that, in this system, we can shift the 
 NS  charge of the D-string  by a multiple of the 
 D-string charge. Undoing the T and S dualities we can get 
 again a system with 3-brane charges $(123)$ and $ (124)$
only. This is T-dual to a $(02)$ system, which we have shown before
 to be U-dual to pure zero branes. 
  As in the $T^2$ discussion, we can map to zero branes 
using only the $O(4,4,Z)$ T-duality group by 
using  theta-shifts and factorized T-dualities.

\subsec{ Gauge theory calculation }

For the maximally supersymmetric solutions, 
  $n \wedge n = 0 mod (N)$. We want
to show that if $gcd (n_{\mu \nu }, N ) = p$ 
then the surviving  zero modes  are 
$U(p)$  flat connections. 
 We review Van Baal's construction \vbal\ of the 
$\Omega_{\mu}$ for arbitrary $N$, and 
$n_{\mu \nu} $ satisfying $n\wedge n = 0 (mod N) $. 
  We use the $SL(4,Z)$ symmetry to 
set all the $n_{ij}=0$ except for $n_{12}, n_{34}, n_{14}$. 
 The case where $n\wedge n = 0$ is simple, since $n_{12}$ or $n_{34}$
is zero. 
For these fluxes we can easily solve for the 
$s_{\mu},t_{\mu}$ as defined in 
\sts. The important property of the resulting $s_{\mu}$ 
$t_{\mu}$ is that only one $t_{\mu}$ is $1$ and the other
$t_{\mu}$ 
are zero, while the $s$'s  satisfy 
$gcd (s_{\mu},N)= gcd (n_{\mu \nu},N)$.

In the general case one decomposes 
\eqn\decN{ 
N= p_{1}^{e_1} p_{2}^{e_2} ... p_l^{e_l}, } 
where 
$p_{i}$ are distinct primes. 
 Correspondingly we decompose the fluxes as 
\eqn\decn{\eqalign{  
& n_{\mu \nu}  = \sum_{i=1}^{l} { N\over {N_i}}n_{\mu \nu}^{(i)} \cr 
& {N\over {N_1}} n_{\mu \nu}^{(1)} + {N \over N_2 }
              n_{\mu \nu}^{(2)}
   + \cdots  {N\over {N_l}} n_{\mu \nu}^{(l)}  \cr } } 
And we decompose $\Omega_{\mu}  = \Omega_{\mu}^{(1)} 
               \otimes \Omega_{\mu}^{(2)} \cdots \otimes 
                \Omega_{\mu}^{(l)} $. 
Each $\Omega^{(i)}$ is of the form $Q_{(i)}$ or $P_{(i)}^{s_i}$. 
Decomposing the gauge field according to the 
same tensor product structure, the surviving 
 gauge connection in each sector is a diagonal  
$U(p_i)$ connection where $p_i$ is the 
$gcd(  n_{\mu \nu }^{(i)}, N_i )$.    The surviving 
flat connections are therefore given by diagonal 
$U(\prod p_i )$ matrices. But by the Lemma  below, 
this is just $U(p)$ flat connections where 
$p= \prod p_i$.

{ \it Lemma }
\eqn\gcdn{ 
 \prod_i  gcd (n_{\mu \nu }^{(i)}, N_i ) = gcd (n_{\mu \nu }, N) }

We first prove that if $\lambda_1 \cdots \lambda_l$ divide 
the pairs $( n_{\mu \nu }^{(1)}, N_1 ), \cdots ( n_{\mu \nu }^{(l)}, N_l )$
then they divide $ (n_{\mu\nu}, N) $. This follows immediately 
from the decompositions   \decN\ and \decn .  
Suppose $\lambda_i = gcd (N_i,n_{\mu \nu}^{(i)})$, 
then $ {n_{\mu \nu} \over {\prod {\lambda_i}}}$
is a sum of terms, each of which is an integer. 
 For example, the first term is 
\eqn\sum{  
{ n_{\mu \nu }^{(1)}\over {\lambda_1 }} 
{N_2 \over {\lambda_2}}\cdots {N_l \over {\lambda_l}} }
which is clearly an integer. 

The converse can be established as follows. 
Let $\lambda = gcd(n_{\mu\nu}, N) $, and
 ${n_{\mu \nu} \over {\lambda}} = q_{\mu \nu}$, 
where $q_{\mu \nu}$ is an integer.
 Given the decomposition 
of $N$ into distinct primes, there must be a
corresponding decomposition 
of $\lambda$ into $\lambda_1 \cdots \lambda_l$, 
where $\lambda_i$ is a power of $p_i$. 
Now let ${ N_i \over {\lambda_i} } = p_i^{k_i}$. Then, 
\eqn\div{\eqalign{  
& q_{\mu\nu} =
{n_{\mu\nu}^{(1)} \over {\lambda_1}} 
{N_2\over{ \lambda_2}} {N_3\over{ \lambda_3}} \cdots 
{N_l\over{ \lambda_l}} + {N_1\over{ \lambda_1}}
 {n_{\mu\nu}^{(2)} \over {\lambda_2}}{N_3\over{ \lambda_3}}\cdots   
  {N_l\over{ \lambda_l}} \cr 
& +  \cdots  + {N_1\over{ \lambda_1} }
{N_2\over{ \lambda_2}}
\dots {n_{\mu\nu}^{(l)} \over {\lambda_l}}
\cr  }}
We want to show that each term is separately an integer, 
i.e for each $i$,  ${n_{\mu\nu}^{(i)}\over {\lambda_i}}$ is 
an integer. Suppose the contrary, e.g that $n_{\mu\nu}^{(1)}$
is not divisible by $\lambda_1$. Then the ratio 
${n_{\mu\nu}^{(1)} \over {\lambda_1}}$ is of the form 
${r_1\over {p_1^{m_1}}}$ for some integer $r_1$ satisfying 
$gcd(r_1, p_1)=1$, and some integer $m_1$. 
Starting from \div\ multiply both sides by 
$\prod p_i^{m_i}$. The LHS is  now divisible by $p_1 \dots p_l$. 
 The first term on the RHS is of the form
$r_1 p_2^{*}\dots p_l^{*}$, where the exponents are all 
integers.   The remaining terms are all divisible by $p_1$. 
The first term on the RHS is not, so the RHS is not divisible 
by $p_1$. This is a contradiction. 
This establishes Lemma 1.

\newsec{Moduli spaces of torons on $T^3$}

   Supersymmetric constant field 
strength solutions exist on odd-dimensional tori, 
 which  are proportional to the 
identity in the Lie algebra of $U(N)$.
The moduli spaces of flat connections associated with such configurations 
can also be deduced from dualities. For concreteness we will describe 
$d=3$. Take  $U(N)$ Yang Mills on $T^3 \times R$, 
which can be thought as worldvolume theory
 of $ N$ 3-branes wrapping the $T^3$. Consider the 
sector with fluxes $n_{12}, n_{23}, n_{13}$. 
 This corresponds to moduli spaces which
are relevant to bound states of 3-branes with  
D-strings. Applying a T-duality along the 
$1$ direction, we get $N$  units of  2-brane
charge  along 
$(23)$ plane, 
$n_{13}$ units of two-brane charge along 
$(12)$ plane, 
$n_{12}$ units of two-brane charge
along the $(13)$ plane, 
and $n_{23}$ units of zero brane charge. 
 We can apply an $SL(3,Z)$ transformation to  get 
the  two-brane charge to lie 
entirely along the $(12)$ plane.  
 The magnitude of the two-brane charge will 
 be the gcd of $N$, $n_{12}$ and $n_{13}$.  
Now by the dualities discussed before 
 we can shift the charges of this system until 
 it is purely 0-brane, and the number of zero branes
is the gcd of the two-brane and the zero brane 
charge we start with. This means that 
the moduli space of supersymmetric  solutions 
in the sector of $U(N)$ gauge theory with fluxes 
$n_{12}$, $n_{23}$, $n_{13}$ is the symmetric 
product $(T^3)^p/S_p$ where $p$ is the gcd of the 
rank and fluxes. 
This can be checked directly in gauge theory 
using the methods described before. 
Similar systems have been discussed in \hv\ 
 where the symmetric dependence 
of the moduli space on the ranks and fluxes finds 
an interpretation in the context of Matrix theory 
at finite $N$ \ls. 

\newsec{  Conclusion and Comments. } 

We have shown that the moduli space of 
torons  in  $U(N)$ Yang Mills theory 
 with  $16$ supersymmetries is equal 
 to the moduli space of $p$ zero branes,  
where $p$ is the greatest common denominator 
of the  Chern classes and rank $N$.
This has been  motivated  by using U-duality. 
In the zero brane description the 
moduli space,   which is a symmetric product, acquires the 
interpretation of positions of $p$ identical particles. 
 We considered 
several dimensions, namely $2$, $3$ and $4$. 
It seems likely that the dependence 
of the moduli space on the common factor of 
 ranks and fluxes  
should be true for higher dimensions, since the combinatorics
leading to it does not seem too dimension dependent (but we 
have not
studied the problem in detail for arbitrary dimensions). 
 Up to 
$10$ dimensions one can imagine embedding 
in some appropriate brane theory and interpreting 
along the lines above. For higher dimensions 
an intriguing possibility is that there might be
 a physical interpretation
analogous to that discussed here, perhaps
in relation to \nish\ or to the speculations
regarding higher dimensions in  \mm.

The moduli space is important in determining 
the spectrum of bound states \bsv\hamo.  
 The counting of one-particle  states
does not depend on the the common factor $p$. 
However when $p$ is greater than one, there can be 
multi-particle states, whose 
multiplicity is related to the  
number of partitions of $p$.  

In the course of describing the dualities 
that map  systems related to torons 
to systems of zero branes 
 we found,  in section 2,  that  some U-duality symmetries
of M theory on $T^3$ have an interpretation in terms 
of boosts in  eleven dimensions. The duality 
group in question has the form $Sl(2,Z) \times SL(3,Z)$. 
The $SL(3,Z)$ is interpreted geometrically in terms 
of reparametrizations of the torus. Some of the 
$SL(2,Z)$ can also be interpreted geometrically 
by using the eleventh dimension of M theory.  
 We saw, in this discussion,  a simple example  
where  boosts in M theory gave  a bigger symmetry for the 
one-particle
spectrum than for the multi-particle spectrum.  
It would be interesting to see if this is  related to 
the analogous facts in Matrix theory \bfss\ discussed in \hv. 

We end by mentioning an interesting  connection 
 between what we  studied here and \witsma\doug. 
In \witsma\doug\ one found a gauge theory whose vacua 
 were in one-one correspondence with 
 the moduli space of instantons on $R^4$, by 
studying the action of a $D$-brane probe 
 in the presence of a $(D+4)$-brane. Here a similar effect 
 is achieved for the case of the torons preserving 
 $16$ supersymmetries. They get related to vacua of a 
$U(p)$ gauge theory on the torus. The intriguing 
 fact is that in this case the physics allowing the relation 
 between instantons and vacua of gauge theories  
 is T-duality, whereas in the case of \witsma\doug, it is the 
  very  different  physics of probe-source interaction.

 \bigbreak\bigskip\bigskip
\centerline{\bf Acknowledgments}\nobreak
It is a pleasure to thank O. Ganor, R. Gopakumar, J. Harvey, 
G.Moore. M. Li, and H. Verlinde 
 for helpful conversations. This  work was supported 
by NSF Grant PHY96-00258 and DOE Grant DE-FGO2-91-ER40671. 

\listrefs

\end